\documentclass[10pt,a4paper]{article}
\usepackage[utf8x]{inputenc}
\usepackage[left=2cm, right=2cm, top=1.50cm, bottom=1.50cm]{geometry}
\usepackage{amsmath}
\usepackage{amsfonts}
\usepackage{amssymb}
\usepackage{graphicx}
\usepackage{theorem}
\usepackage{subfig}
\usepackage{psfrag}
\usepackage{epstopdf}
\usepackage[hyphens]{url}

\usepackage{url}
\usepackage{colortbl}
\theorembodyfont{ \normalfont }

\usepackage{hyperref}
\newtheorem{theo}{Theorem}

\begin{document}
\begin{flushleft}
Proceedings 	of the {\em IX Encontro dos Alunos e Docentes do Departamento de Engenharia de Computa\c{c}\~{a}o Automa\c{c}\~{a}o At: Campinas $29$ e $30$ de setembro de 2016, SP--Brazi}, available in   \href{https://www.fee.unicamp.br/sites/default/files/departamentos/dca/eadca/eadcaix/artigos/carvalho_et_al.pdf}{[$2016$ EADCA-IX]}.
\end{flushleft}
\vspace{2 mm}

\begin{center}
\textbf{	{\LARGE  The Burst Failure Influence on the $H_\infty$ Norm} } 
\end{center}
\vspace{3 mm}		
\textbf{Leonardo de P. Carvalho, Jonathan M. Palma, Lucas P. Moreira and Alim P. C. Gon\c{c}alves~(Orientador).}

		Departamento de Engenharia de Computa\c{c}\~{a}o e Automa\c{c}\~{a}o Industrial (DCA).    Faculdade de Engenharia El\'{e}trica e de Computa\c{c}\~{a}o (FEEC).  Universidade Estadual de Campinas (Unicamp).   		Caixa Postal 6101, 13083-970 -- Campinas, SP, Brasil. 	{\tt  email \{jpalmao,lcarvalho,lporrelli,alimped\}@dca.fee.unicamp.br}\\

\textbf{ \em Abstract} \textbf{-- In this work, we present  an analysis of the Burst failure effect in the $H_\infty$ norm. We present a procedure to perform an analysis between different Markov Chain models and a numerical example. In the numerical example the results obtained pointed out that the burst failure effect in the performance does not exceed $6.3\%$. However, this work is an introduction for a wider  and more extensive analysis in this subject.}\\

\textbf{{\em keywords:  $H_\infty$ norm, Networked Control Systems, Burst-Failure}}

\section{Introduction}
In the last decade, the Networked Control System theory has been heavily studied, primarily because it makes possible to design control solutions, which consider the problems inherent to the network,\cite{CostaFragoso},\cite{Hespanha}. One of the most problematic characteristic in a network is that it will always have some sort of packet loss. Usually, this packet loss is solved  re-transmitting the same information until it reaches its destination. But depending on the amount of re-transmission, this can lead to another problem, the delay,\cite{Ovsthus}. The packet loss has a peculiar behavior, when the network fails the chance of the network fails again in the next instant $k$ is high, this leads to several consecutive failures. This characteristic is named burst failure,\cite{Marcondes}.

Usually, when designing a controller via Networked Control System(NCS) the only information about the network is the packet loss rate (PLR). If the controller is designed with only the PLR information, the network behavior that the controller is expecting from the network is a Bernoulli Process, and like we explained before, the network has a more complex behavior. Moreover, the Bernoulli process does not contemplate the burst failure like others Markov Chain models, (e.g. Gilbert, Gilbert Eliot, McCullough). When this approximation is made some information about the network behavior is lost. It is well known that depending on the system dynamic the $H_\infty$ norm is very sensitive to the packet loss variation. In this situation, the Bernoulli approximation may influence the performance presented by the controller designed using only packet loss rate as information that represents the entire network behavior.

In this work we study the influence when the controllers are designed using the Bernoulli process as a network model compared with controllers designed using other Markov Chains that contemplate the burst failure, like the Gilbert model and the Gilbert-Eliot model. We are using the $H_\infty$ norm as a performance measurement, the $H_\infty$ norm is commonly used because it is a system characteristic that represents the robustness of the system.

At the end of the work we present an example to show the comparison between the $H_\infty$ norm, using the Bernoulli model  and the worst case scenario using the Gilbert model. 

\section{Theoretical Background}
In this section, we explain the necessary theoretical background to understand and replicate the results exposed in this work.
\subsection{Discrete-Time Markov Chains}
A Markov Chain is a random process that the next state  depends only on the present state. The Markov chains can be classified as infinite or finite chain, in the infinite chain the set of  states is unlimited, and otherwise for the finite chain. The transition between mode follows the probabilities contained in the transition probability matrix, an example of this matrix is presented below,
\begin{eqnarray}
{\color{blue}	\mathbb{P}=\begin{bmatrix}
		1- p &   p \\
		 q & 1-q
	\end{bmatrix}}
\end{eqnarray} 
It is possible to calculate the n-step transition probability using $P^n$,\cite{Leon}. 

\subsubsection{Steady State}
Some Markov Chain after a long time running, $n \rightarrow \infty$, it settles into a stationary state. In the Steady State the matrix $P$ all the rows are equal to the same probability mass function.

\subsection{Markov Jump Linear System}\label{sec:MJLS}
A Markov Jump Linear System (MJLS) can be defined as a class of switched system , where the permutation is a stochastic process. The permutation is made between modes. The modes are subsystems with its own particularities. The permutations are made according to a Markov Chain. A general mathematical representation is described by Eq.\eqref{eq:sistema},
\begin{equation}\label{eq:sistema}
	\mathcal{G}  :
	\begin{cases}
		x(k+1)&=A(\theta_k)x(k)+J(\theta_k)w(k),\\
	{\color{blue} z(k)} &=C_y(\theta_k)x(k)+E_y(\theta_k)w(k),              
	\end{cases}
\end{equation}
where $x(k) \in \mathbb{R}^r$ is the state vector, $w(k) \in \mathbb{R}^m$ is the exogenous input (i.e., the signal that represents the disturbance or noise). The $y(k)$ vector denotes the estimated output. The variable $\theta_k$ is a random variable and assume values on the finite set $\mathbb{K} = \{1,2,\cdots,N \}$, each value representing a specific mode and, in every instant $k$, the mode may or may not change. These transitions between modes occurs according to a Markov Chain. Probability is given by $p_{ij} = Prob(\theta_{k+1} = j|\theta_k=i), p_{ij}>0$ and $\sum_{i \in \mathbb{K}}^n p_{ij}=1 , \ \forall \  i,j \in \mathbb{K}$. The transition matrix is represented by $\mathcal{P} = [p_{ij}]$. To simplify the notation, for now on, whenever $\theta_k=i$, we write  $A(\theta_k) = A_i$. More information about Markovian Linear Jump Systems can be found in \cite{CostaFragoso}.

\subsection{Markovian $H_\infty$ norm}
The $H_\infty$  norm can be interpreted as the maximum cost threshold for the worst disturbance influence on the output. A formal definition is expressed in Equation \ref{equ-markovian-infinity-norm}:
\begin{equation}
	\| \mathcal{G} \|^2_\infty = \sup_{0 \neq w \in \mathcal{L}^2,\ \theta_0 \in \mathcal{K}}  \frac{\|z	{\color{blue}(k)}  \|^2_2}{\|w	{\color{blue}(k)} \|^2_2}\label{equ-markovian-infinity-norm}
\end{equation}

a more detailed explanation about Markovian $H_\infty$ norm can be found in \cite{Geromel,Zhou}.

In order to calculate the $H_\infty$ norm, we use the following theorem,

\begin{theo}
	Given the system \eqref{eq:sistema} stable by the second moment, the $H_\infty$ norm from the relation between the output $z$ and the input $w$ is bounded by $\| \mathcal{G} \|^2_\infty < \gamma$ if, and only if, there exist symmetric matrices $P_i=P_i' >0 \ \forall \ i \in \mathbb{K}$, $P_{pi}=\sum_{j\in\mathbb{K}} p_{ij}P_j$ and $\gamma > 0$,
	\begin{equation}
		\begin{bmatrix}
			P_i & \bullet & \bullet & \bullet \\
			0 & \gamma I & \bullet & \bullet \\
			P_{pi} A_i & P_{pi} J_i & P_{pi} & \bullet \\
			C_{zi} & E_{zi} & 0 & I
		\end{bmatrix}>0,
	\end{equation}
	if a feasible solution for all $i \in \mathbb{K}$ is found and the system is weakly controllable the norm is given by 
	$\| \mathcal{G} \|_\infty < \sqrt{\gamma}$.
\end{theo}
The prove of this theorem is found in \cite{Geromel}.
\subsection{Probabilistic Models}
\begin{figure}[th!]
	\centering
	\subfloat[Gilbert model]{\includegraphics[scale=0.19]{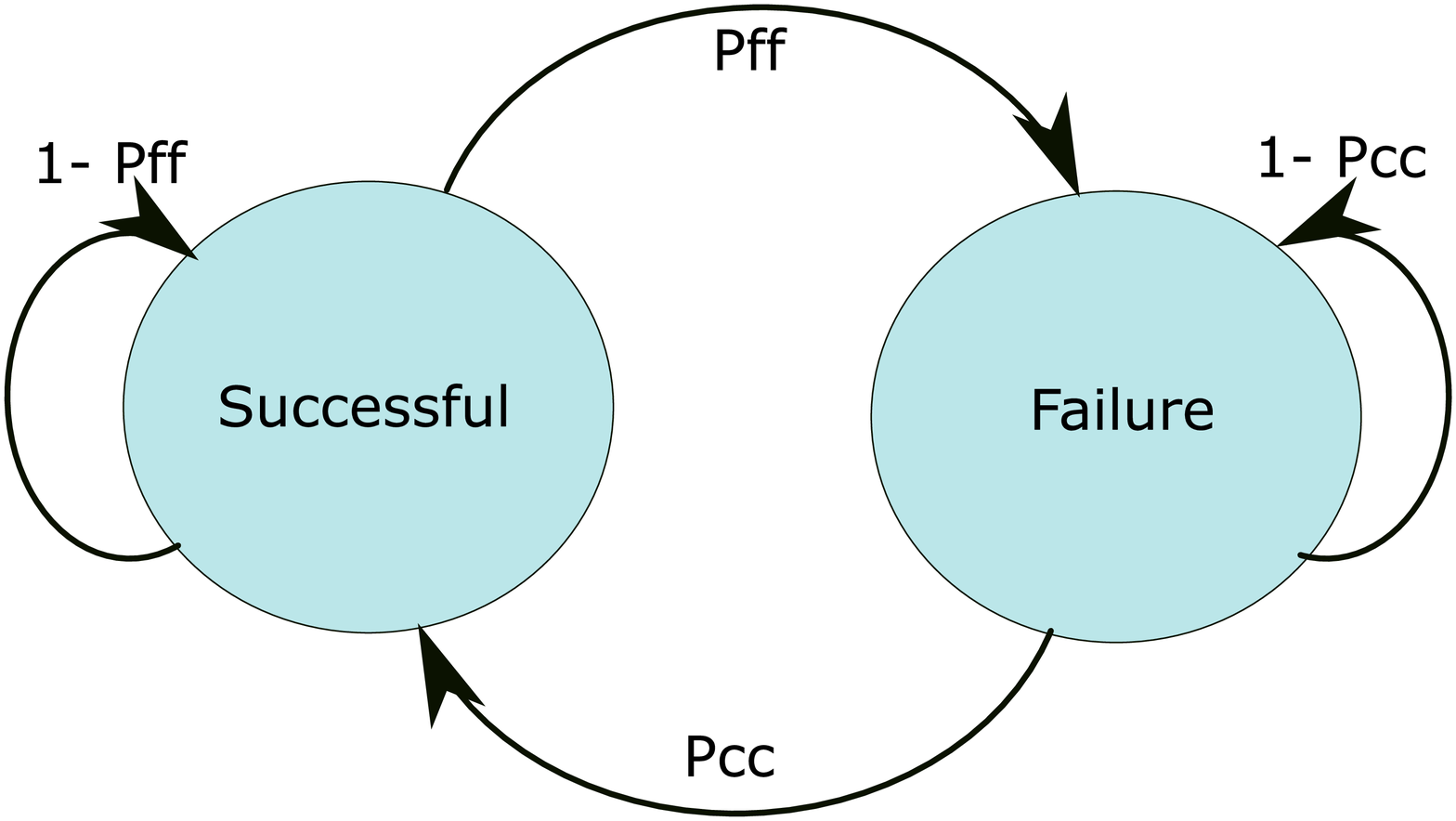}}\label{fig:gildese}
	\subfloat[Bernoulli model]{\includegraphics[scale=0.19]{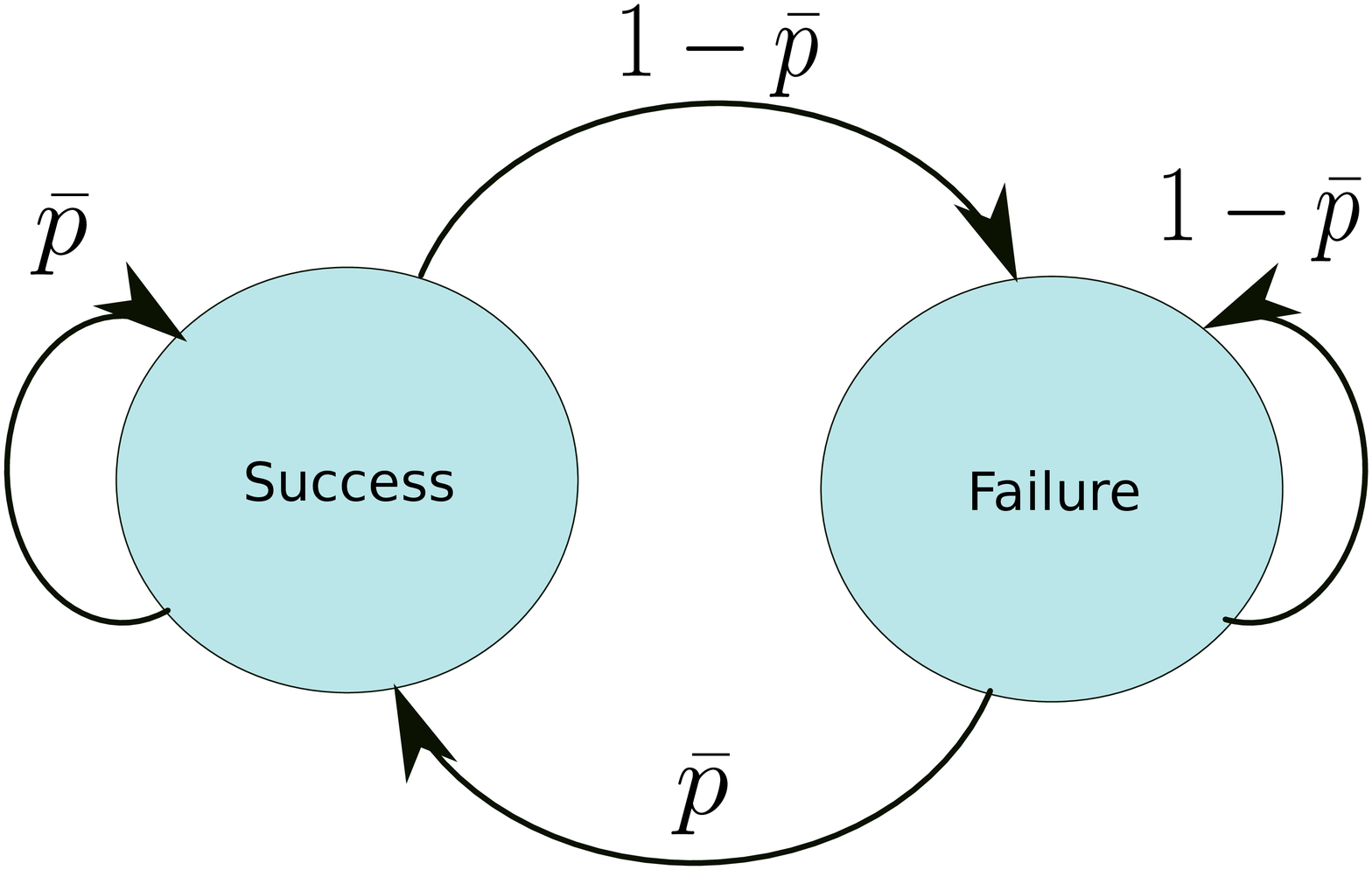}}\label{fig:berdese}
	\caption{Graphic representation for both models, {\color{blue} with $q=P_{cc}$ and $p=P_{ff}$}}\label{fig:deses}
\end{figure}

In this work we use two types of probabilistic models: the Bernoulli model and the Gilbert model. The Bernoulli model is a special case for the  Gilbert model. The network model is given by a Gilbert model which possesses similar characteristics to burst failures present in real networks. Figure \ref{fig:deses} presents a graphical representation for both modes.
The probability matrix, $P_{gil}$, for the Gilbert model and , $P_{ber}$, for the Bernoulli model (see \cite{Gilbert} for a complete description for the Gilbert model and \cite{Leon} for the Bernoulli model).

\begin{align}
	\mathbb{P}_{gil}=
	{\color{blue} \begin{bmatrix}
		1- p &   p \\
		q & 1-q
		\end{bmatrix}},
	& \ \
	\mathbb{P}_{ber}=
	\begin{bmatrix}
		\bar{p} & 1-\bar{p} \\
		\bar{p} & 1-\bar{p}
	\end{bmatrix}
\end{align}
\section{Procedure to analyze the burst failure effect on the $H_\infty$}
\subsection{Steady State}
\begin{figure}[ht!]
	\centering
	\includegraphics[scale=0.22]{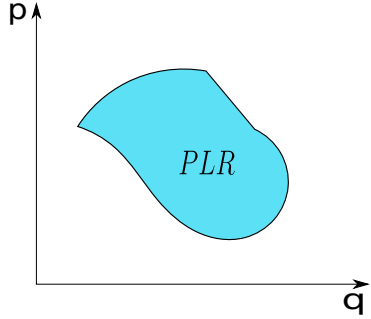}
	\caption{Graphic representation for region with the same PLR}\label{fig:subspace}
\end{figure}

In order to make a fair comparison between the model, we need to compare the models in situation that both have the same probability mass function. For the Bernoulli model with a specific packet loss rate value, the Gilbert mode has a region of values of $p$ and $q$ with the same packet loss rate in the steady state. The Fig. \ref{fig:subspace} shows a representation of this situation.
In order to describe this region, we need to make the following procedure, first, we need to consider that,
\begin{equation}\label{eq:pnp}
	P_{gil}^n = P_{ber}
\end{equation}
This consideration means that the steady state matrix for the Gilbert mode is equal to the matrix that represents the Bernoulli model. (The second step is to find the exponential $P_{gil}^n$. Making the diagonalization method in $P_{gil}$, we obtained the following matrix.\cite{Leon})
\begin{eqnarray}
	P_{gil} =& E \Lambda E^{-1} \\
	=&\frac{1}{p + q} 
		\begin{bmatrix}
			1 & p \\
			1 & -q
		\end{bmatrix}
		\begin{bmatrix}
			1 & 0 \\
			0 & 1 - p - q
		\end{bmatrix}
		\begin{bmatrix}
			q & p \\
			1 & -1
	\end{bmatrix}
\end{eqnarray}
With $P_{gil}$ diagonalized, we are able to calculate $P_{gil}^n$ as presented below,
\begin{eqnarray}
	P_{gil}^n =& (E \Lambda E^{-1})^n \\
	=& (E \Lambda E^{-1}) \times (E \Lambda E^{-1})  \times \cdots \times (E \Lambda E^{-1})\\
	=& E \Lambda^n E^{-1}\\
	=& \frac{1}{p + q} 
		\begin{bmatrix}
			1 & p \\
			1 & -q
		\end{bmatrix}
		\begin{bmatrix}
			1 & 0 \\
			0 & (1 - p - q)^n
		\end{bmatrix}
		\begin{bmatrix}
			q & p \\
			1 & -1
	\end{bmatrix} \\
	=& \begin{bmatrix}
			\frac{q}{p + q} & \frac{p}{p + q} \\
			\frac{q}{p + q} & \frac{p}{p + q}
		\end{bmatrix}+
		\frac{(1-p - q)^n}{p+ q}
		\begin{bmatrix}
			p & p \\
			q & q 
	\end{bmatrix}
\end{eqnarray}
as long as $|1- p - q\ |< 1$ when $n\rightarrow \infty$ the second term goes to zero so we can say that,
\begin{eqnarray}
	P_{gil}^n = \begin{bmatrix}
		\frac{q}{p + q} & \frac{p}{p + q} \\
		\frac{q}{p + q} & \frac{p}{p + q}
	\end{bmatrix}
\end{eqnarray}
Now we can find the region for all pairs $p$ and $q$ that have the same PLR as a Bernoulli process. This region is formed by two equations obtained using the equation \eqref{eq:pnp},those equations are presented below,
\begin{eqnarray}
	PLR = \frac{q}{p + q} \\
	(1-PLR) = \frac{p}{p + q}
\end{eqnarray}
In possession of these equations we can find the region for a specific $PLR$.
\section{Numerical example}
In this section we present the preliminary results obtained for experiment.
\subsection{Dynamical system}
\begin{figure}[th]
	\centering
	\includegraphics[scale=0.7]{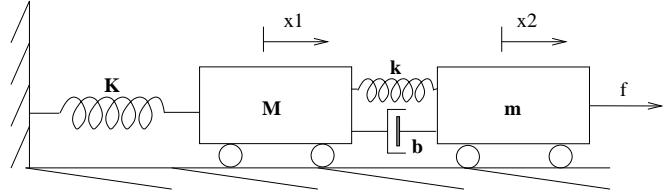}
	\caption{Studied physical system.}\label{fig:carro}
\end{figure}

The model used in this example consists of a mass spring damp system composed by two cars with masses $M$ and $m$ interconnected by a spring and dampening setting. The car whose mass is $M$ is linked to a wall by a spring. Fig. \ref{fig:carro} shows the studied physical system. The continuous time
equations are easily obtained by classical modeling. Thus, taking into account the respective parameters, the resulting matrices of the continuous state space form,
\begin{align}\label{eq:sysnum}
	\left[\begin{array}{c|c}
			A & J \\ \hline
			C_{y1} & E_{y1} \\ \hline
			C_{y2} & E_{y2}
		\end{array}\right] &= 
		\left[\begin{array}{cccc|cccc}
			0 & 0 & 1 & 0 & 0.5 & 0 & 0 & 0\\
			0 & 0 & 0 & 1 & 0 & 0.5 & 0 & 0\\
			0 & 26.29 & -15.96 & -0.02 & 0 & 0 & 0.5 & 0\\
			0 & 68.52 & -15.25 & -0.04 & 0 & 0 & 0 & 0.5 \\ \hline
			1 & 0 & 0 & 0 & 0.05 & 0 & 0 & 0\\
			0 & 1 & 0 & 0 & 0 & 0.05 & 0 & 0 \\ \hline
			0 & 0 & 0 & 0 & 0 & 0 & 0 & 0 \\
			0 & 0 & 0 & 0 & 0 & 0 & 0 & 0\\ 
		\end{array}\right].
\end{align}
The only measured state is the first car position, $x_2$, with a $0.5\%$ noise in this measurement. This example has two mode the first mode represents the situation when the measurement signal is properly received, represented as $C_{y1}$ and $E_{y1}$, respectively. The second mode is when the signal is lost during the transmission, represented as $C_{y2}$ and $E_{y2}$. The discretization time used was $T=0.01\mbox{ s}$.
\subsection{Finding the region}
The first results are regions where the PLR after $n$ interactions, with $n \rightarrow \infty$, is the same as the PLR for the Bernoulli process. We found the regions that represent the following values of PLR $[0.1\ 0.2 \ 0.3 \ 0.4 \ 0.5 \ 0.6 \ 0.7 \ 0.8\ 0.9]$. This graphic is presented below,

\begin{figure}[th]
	\centering
	\includegraphics[scale=0.8]{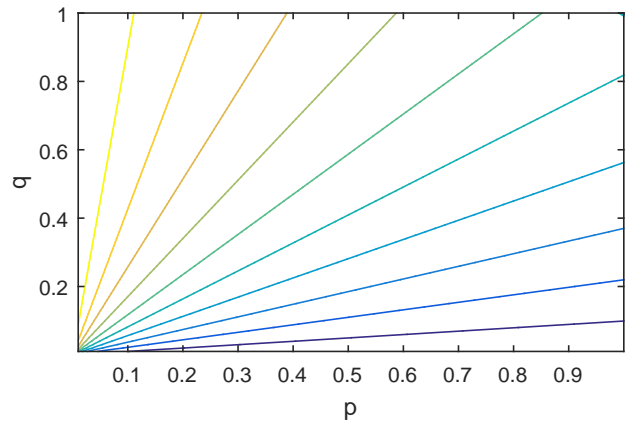}\label{fig:subspace_results}
	\caption{{\color{blue}Graphic showing the region where the PLR is equal after $n$ iterations}}
\end{figure} 
The precision used to find these regions was $0.0001$. It is possible to observe that the regions for each PLR has different sizes and does not have any intersection between these regions. These characteristics are expected because a specific pair $[p \ q]$ has only one value of PLR.
\subsection{Obtaining the worst $H_\infty$ norm in the region}
The second step is find the pair $p,q$ with the highest $H_\infty$ norm, the worst case, and compare with $H_\infty$ norm obtained with $P_{ber}$. The results for each value of PLR is presented in the table below.
\begin{table}[th]
	\centering
	\label{tab:table1}
	 \begin{tabular}{|c|c|r|r|}\hline
			PLR & $H_\infty$GIL & $H_\infty$BER & ERR\\
			\hline
			0.1 & 1.1435 & 1.1315 & 0.1$\%$\\
			0.2 & 1.5430 & 1.5346 & 0.5$\%$\\
			0.3 & 1.8511 & 1.8268 & 1.3$\%$\\
			0.4 & 2.1033 & 2.0893 & 0.67$\%$\\
			0.5 & 2.3317 & 2.3264 & 0.22$\%$\\
			0.6 & 2.5923 & 2.5362 & 2.21$\%$\\
			0.7 & 2.8576 & 2.7425 & 4.19$\%$\\
			0.8 & 3.1023 & 2.9183 & 6.30$\%$\\
			0.9 & 3.2198 & 3.1131 & 3.42$\%$\\ \hline
	\end{tabular}\caption{Table of comparison.}
\end{table} 
It is possible to observe that when PLR increases the Burst failure relevance also increases for this specific example. However, the discrepancy found between the $H_\infty$ norm does not exceed $6.30\%$, this value shown that the difference in the performance for this specific example is not relevant. But, it is also important to point out that the worst noise for the Bernoulli model is not necessarily the same worst noise for the Gilbert model, and in a implementation the difference in the performance can improve or decrease depending on the type of noise that the system will be exposed.

\section{Conclusion}
In this work we presented the preliminary analyses made on the effects that the Burst failure has in the $H_\infty$ norm. We introduced a procedure to make a fair comparison between different Markov chain models and presented an example of one Markov chain model, the Gilbert Model. The results obtained in this first attempt to make this analysis, where  the difference in performance does not exceed $6.3\%$, apparently, showing that the burst failure has not a relevant in $H_\infty$ norm, however, it is necessary to perform more test in order to verify such assumption. Like we explained in the previous section, the worst noise in for the Bernoulli Model is not necessarily equal to the worst noise in the Gilbert Model, with that information in mind, an important analysis would be the a Temporal Monte Carlo simulation with noises that are relevant to the system in the example. Another important analysis would be the analyses of other Markov Chain models such the Gilbert-Eliot and McCullough models, these models are better network representations than the Gilbert model and they would add more veracity to the experiment. This first analyses was important because now we have the information about the region characteristic, now that we know that it is a convex region. With that information in hands, we will be able to create solutions that only with the PLR information we will be capable of design a controller that  consider the burst failure.   


%

\begin{thebibliography}{99}
	\bibitem{CostaFragoso} Costa, O.L.V., Fragoso, M.D., Marques, R.P.: Discrete-time Markov jump linear systems. Springer Science and Business Media, 2006.
	
	\bibitem{Hespanha} Hespanha, Joao P., Payam Naghshtabrizi, and Yonggang Xu. "A survey of recent results in networked control systems." PROCEEDINGS-IEEE 95.1 (2007): 138.
	
	\bibitem{Ovsthus} Ovsthus, Knut, and Lars M. Kristensen. "An industrial perspective on wireless sensor networks a survey of requirements, protocols, and challenges." IEEE communications surveys and tutorials 16.3 (2014): 1391-1412.
	
	\bibitem{Marcondes} Marcondes, G. A. B. Modelos discretos para análise de ocorrências de erros em redes sem fio. Diss. M. Sc. Thesis, Inatel, Brazil, 2005 (in portuguese), 2005.
	
	\bibitem{Leon} Leon-Garcia, Alberto, and Alberto. Leon-Garcia. Probability, statistics, and random processes for electrical engineering. Upper Saddle River, NJ: Pearson/Prentice Hall, 2008.
	
	\bibitem{Geromel} Geromel, Jos\'{e} C., and R. H. Korogui. "Controle Linear de Sistemas Din\^{a}micos." Editora Blucher, S\~{a}o Paulo, SP (2011).
	
	\bibitem{Zhou} Zhou, Kemin, John Comstock Doyle, and Keith Glover. Robust and optimal control. Vol. 40. New Jersey: Prentice hall, 1996.
	
	\bibitem{Gilbert} Gilbert, Edgar N. "Capacity of a Burst Noise Channel.", Bell system technical journal 39.5, 1960, pg. 1253 1265.
\end{thebibliography}
\end{document}